\documentclass[10pt, a4paper, twocolumn, twoside, amssymb]{article} 
\pdfoutput=1

\usepackage[english]{babel} 

\usepackage{microtype} 

\usepackage{amsfonts,amsthm} 
\usepackage{amssymb} 
\usepackage{isomath} 

\usepackage[svgnames]{xcolor} 

\usepackage[small, labelfont=bf, up]{caption} 

\usepackage{booktabs} 

\usepackage{lastpage} 

\usepackage{graphicx} 

\usepackage{enumitem} 
\setlist{noitemsep} 

\usepackage{sectsty} 
\allsectionsfont{\usefont{OT1}{phv}{b}{n}} 

\usepackage{geometry} 

\usepackage[T1]{fontenc} 
\usepackage[utf8]{inputenc} 

\usepackage{XCharter} 

\usepackage{hyperref}

\usepackage{journals}

\usepackage{fancyhdr} 
\newcommand{\shorttitle}[1]{\fancyhead[CE]{\textsl{#1}}}
\newcommand{\shortauthors}[1]{\fancyhead[CO]{\textsl{#1}}}

\fancypagestyle{firstpage}{ 
	\fancyhf{}
        \lhead{\vspace*{0.0em} \textit{\footnotesize 21$^{\rm st}$ European Workshop on White
            Dwarfs \\ Castanheira, Vanderbosch, Montgomery, eds. \\[-0.2em]
          July 23--27, 2018, Austin, Texas, USA}}
}

\date{}

\newcommand{\authorstyle}[1]{{\large\usefont{OT1}{phv}{b}{n}\color{DarkRed}#1}} 

\newcommand{\institution}[1]{{\footnotesize\usefont{OT1}{phv}{m}{sl}\color{Black}#1}} 

\usepackage{titling} 

\newcommand{\HorRule}{\color{DarkGoldenrod}\rule{\linewidth}{1pt}} 

\pretitle{
	\vspace{-30pt} 
	\HorRule\vspace{10pt} 
	\fontsize{16}{12}\usefont{OT1}{phv}{b}{n}\selectfont 
	\color{DarkRed} 
}

\posttitle{\par\vskip 10pt} 

\preauthor{} 

\postauthor{ 
	\vspace{0pt} 
	{\par\HorRule} 
	\vspace{-10pt} 
}

\newcommand{\newabstract}[1]{
    {\section*{Abstract}
    \bfseries #1}
  }

\usepackage[authoryear]{natbib}
\newcommand\teff{\ensuremath{T_{\rm eff}}}
\newcommand{\bvfreq}{Brunt-V\"ais\"al\"a frequency}
\newcommand{\brv}{Brunt-V\"ais\"al\"a}
\newcommand\pcross{\ensuremath{P_{\rm cross}}}

\title{Limits on Mode Coherence Due to a Non-static Convection Zone} 

\shorttitle{Limits on Mode Coherence Due to a Non-static Convection
  Zone} 

\shortauthors{Montgomery, Hermes, and Winget} 

\author{
        \authorstyle{M.~H.~Montgomery,$^1$ J.~J.~Hermes,$^2$ and
          D.~E. Winget$^1$}
	\newline\newline 
	$^1$\institution{University of Texas and McDonald Observatory, Austin, TX, USA; 
          mikemon@astro.as.utexas.edu, dew@astro.as.utexas.edu}\\ 
	$^2$\institution{Hubble Fellow, Department of Physics and Astronomy, University of 
North Carolina, Chapel Hill, NC, USA; jjhermes@unc.edu}
}

\begin{document}

\renewcommand{\floatpagefraction}{.8}

\maketitle 

\thispagestyle{firstpage} 


\newabstract{ The standard theory of pulsations deals with the
  frequencies and growth rates of infinitesimal perturbations in a
  stellar model.  Modes which are calculated to be linearly driven
  should increase their amplitudes exponentially with time; the fact
  that nearly constant amplitudes are usually observed is evidence
  that nonlinear mechanisms inhibit the growth of finite amplitude
  pulsations. Models predict that the mass of DAV convection zones is
  very sensitive to temperature (i.e.,
  $\mathbfit{ M_{\rm \bf CZ} \propto T_{\rm \bf eff}^{-90}}$), 
  leading to the
  possibility that even ``small amplitude'' pulsators may experience
  significant nonlinear effects.  In particular, the outer turning
  point of finite-amplitude g-mode pulsations can vary with the local
  surface temperature, producing a reflected wave that is slightly out
  of phase with that required for a standing wave. This can lead to a
  lack of coherence of the mode and a reduction in its global
  amplitude. We compute the size of this effect for specific examples
  and discuss the results in the context of \emph{Kepler} and
  \emph{K2} observations.}


\section{Astrophysical Context}

In the linear, adiabatic theory of stellar pulsations, modes are
considered to be perfectly sinusoidal in time. This results in a
theoretical eigenmode spectrum with arbitrarily thin, delta function
peaks as a function of frequency. In the linear, \emph{non}-adiabatic
theory, modes are allowed to gain or lose energy with their
environment, leading to non-zero growth/damping rates, and these rates
can be interpreted as finite widths of the peaks in their power
spectra. In particular, the widths of modes in solar-like pulsators
can be linked to their computed linear damping rates
\citep[e.g.,][]{Houdek15,Kumar89}.

In the white dwarf regime, we have direct observational evidence that
some modes in pulsating white dwarfs can show a high degree of phase
coherence, in a few cases spanning decades. For instance, in the case
of the DAV G117-B15A, over 40 years of observations have established
that not only is the phase of the 215~s mode coherent over this time
scale, the pulsation period, $P$, is changing (taking into account the
proper-motion effect) at the extremely slow rate of
$\dot{P} \rm = (3.57 \pm 0.82) \times 10^{-15} s\,s^{-1}$
\citep{Kepler05a}. The extreme sensitivity of such $\dot{P}$
measurements in this and other white dwarfs has allowed them to be
used as testbeds for unknown physical processes that could affect
their cooling, such as the emission of hypothetical particles
\citep[e.g.,][]{Isern08,Kim08b,Corsico12a,Corsico16}. Another use of
the observed stability of these modes is searching for planetary
signals in the delayed and advanced light arrival times due to reflex
orbital motion of the white dwarf
\citep{Mullally03,Mullally08,Hermes10,Winget15}; to date, no planets
orbiting WDs have been positively identified with this technique.


While a handful of such coherent modes have been studied in DAVs, no
systematic study of mode coherence has been made from the ground.
This situation has improved greatly with the launch of the
\emph{Kepler} spacecraft. During its original mission and the
follow-on \emph{K2} mission, it has obtained nearly-continuous time
series data, often exceeding 75 d, of a large number of pulsating
white dwarf stars. Recently, \citet{Hermes17a} published comprehensive
data on 27 DAVs studied by \emph{Kepler}. One of their central results
was that longer period modes ($P \gtrsim 800$~s) were observed to have
larger Fourier widths than shorter period modes ($P \lesssim 800$~s),
essentially dividing the modes into two populations; this result even
holds for different modes in the \emph{same} star. We present an
explanation for this phenomenon in terms of the differing propagation
regions of these two classes of modes and show how this could be used
to constrain different models of convection.

\section{The Data}

The extended length of observations ($\gtrsim 75$~d for most stars) in the
\emph{Kepler} and \emph{K2} data sets results in a $1/T$ resolution in
the power spectra of $0.14 \,\, \mu \rm Hz$; this sets the observable
lower limit for the width of a peak in the Fourier transform. For the
first time this enables the measurement of the widths of a large
number of modes in many stars that are above this threshold.

\begin{figure}[t]
  \vspace*{0.5em}
  \centering{\includegraphics[width=\columnwidth]{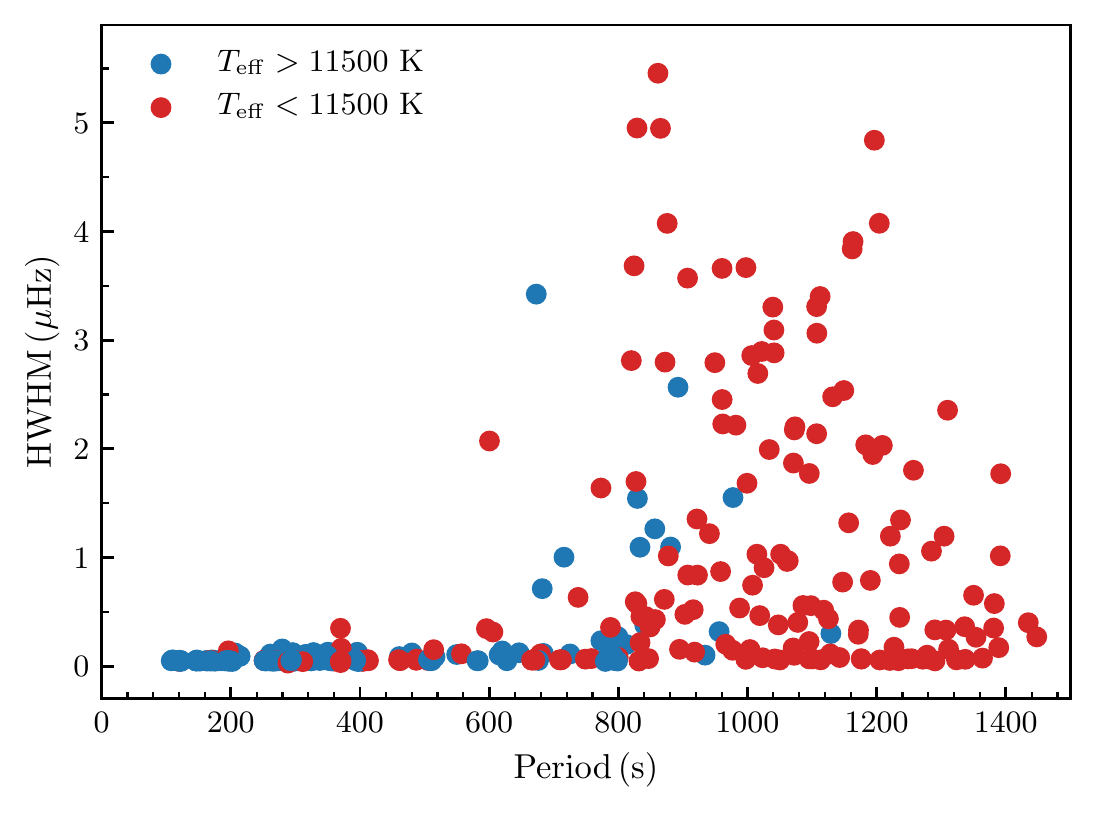}} 
  \caption{
    Observed Fourier width versus period for modes in DAVs observed
    with \emph{Kepler} and \emph{K2}. The blue and red points are for
    modes in stars with $\teff > 11$,500~K and $\teff < 11$,500~K,
    respectively.
    }
\label{hermes}
\end{figure}

\citet{Hermes17a} find that the Fourier width of modes, ($\Gamma$, the
half width at half maximum), is a strong function of the mode period,
$P$. To summarize, they find that 1) modes with $\Gamma > 0.3\,\mu$Hz
have $P \gtrsim 800$~s and 2) modes with $P \lesssim 800$~s have
$\Gamma < 0.3\,\mu$Hz.  This is illustrated in Fig.~\ref{hermes}, in
which we have plotted mode width versus period for all the linearly
independent periods found in the sample of 27 DAVs from
\citet{Hermes17a}. Modes from stars with $\teff > 11$,500~K are shown
as blue points, while those from stars with $\teff < 11$,500~K are
shown as red points. The fact that most of the modes with
$\Gamma > 0.3\,\mu$Hz are from the cooler population is not an
independent piece of information since longer-period modes are known
to be found in the cooler DAVs \citep{Mukadam06b}.

In this paper we introduce a mechanism that should be present for
limiting the growth of mode amplitudes. This mechanism becomes more
important at cooler temperatures and might be relevant to the red edge
of the DAV instability strip.

\begin{figure}[t]
  \centerline{\includegraphics[width=1.0\columnwidth]{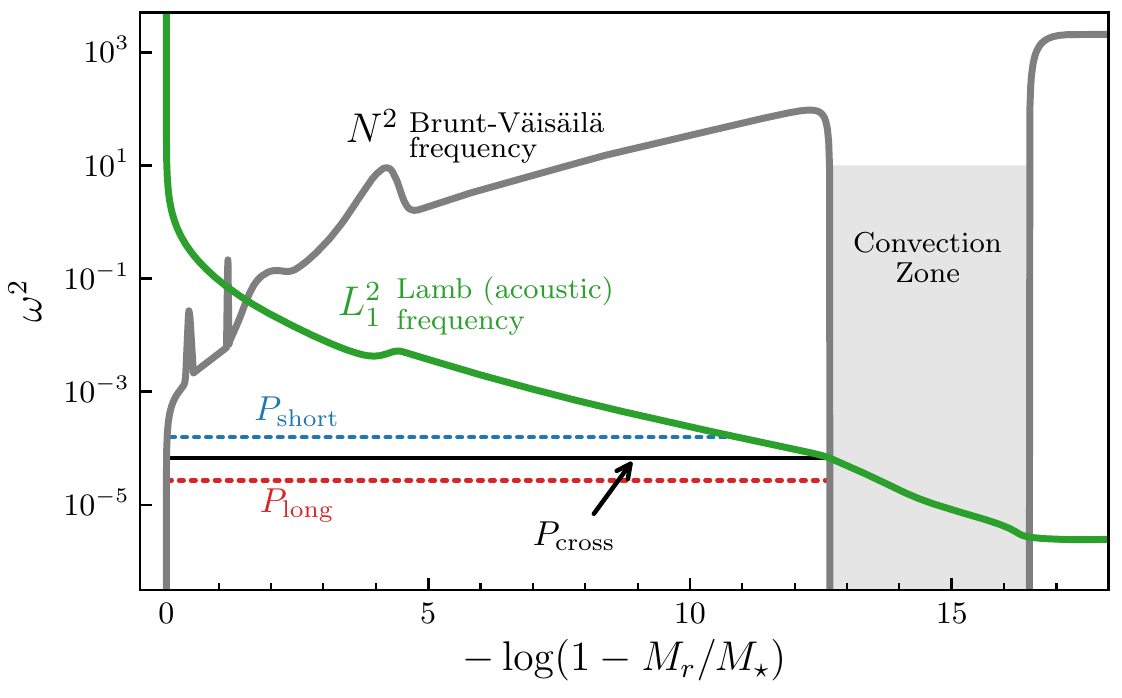}}
  \caption{Schematic propagation diagram of a DAV white dwarf.  A
    g-mode with a ``short'' period would have an outer turning point
    beneath the convection zone (blue dotted line), while one with a
    ``long'' period would have an outer turning point at the
    convection zone boundary (red dotted line). The crossover point
    between these two regimes is given by a mode with
    $P=P_{\rm cross}$ (black horizontal line).  }
  \label{prop1}
\end{figure}

\begin{figure}[b!]
  \centerline{\includegraphics[width=1.0\columnwidth]{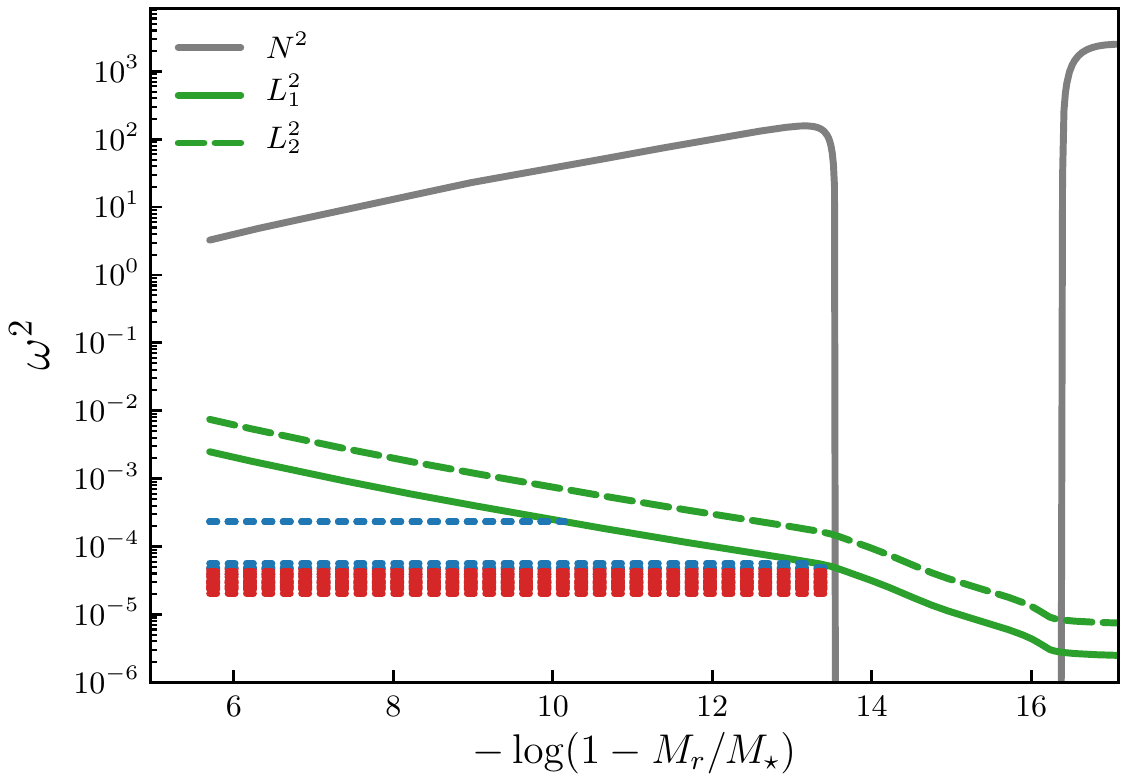}}
  \caption{The outer propagation region for a model with the same
    parameters as EPIC 201806008. The dashed horizontal lines show the
    region of propagation that the observed modes would have in this
    model: the blue dashed lines denote modes observed to have
    $\Gamma < 0.3\, \mu$Hz and the red dashed lines modes with
    $\Gamma > 0.3\, \mu$Hz.}

  \label{epic1}
\end{figure}

\section{The Propagation Region}

The region of propagation of $g$ modes in a star is defined by 
the region in which
$\omega^2 < N^2, L^2_{\ell}$,
where $\omega = 2 \pi/P$ is the angular frequency of the mode,
$N$ is the \brv\ (buoyancy) frequency, and
$L_{\ell}$ is the Lamb (acoustic) frequency. In Fig.~\ref{prop1}, we
show a propagation diagram for a model, computed with MESA
\citep{Paxton11,Paxton13,Paxton15,Paxton18}.  The grey
horizontal line denotes the region of propagation of a hypothetical
mode whose period, \pcross, is the minimum required for it to
propagate to the base of the convection zone.

In Fig.~\ref{epic1}, we show the outer propagation region for a model
with the same parameters as EPIC 201806008 ($\log \,g=8.02$, $\teff =
10910\,$K).  The dashed horizontal lines show the region of
propagation that the observed modes would have in this model, where
the blue dashed lines denote modes observed to have $\Gamma < 0.3\,
\mu$Hz and the red dashed lines modes with $\Gamma > 0.3\,
\mu$Hz.  We note that, by setting $\alpha =
0.42$ for this model, we can divide the modes into two groups: 1)
narrow (blue) modes whose outer turning point is well beneath the
convection zone, and 2) wide (red) modes that propagate all the way to
the base of the convection zone. In other words, all the modes with
$\Gamma >
0.3\,\mu$Hz can be explained as propagating to the base of the
convection zone, whereas all the modes with $\Gamma <
0.3\,\mu$Hz have an outer turning point safely inside this point.
{\bf Our hypothesis is that the long-period modes, through their
  interaction with the convection zone, will have systematically
  larger Fourier mode widths than the short-period modes.} In the
following sections we examine this statement more quantitatively.

\begin{figure}[t!]
  \centerline{\includegraphics[width=1.0\columnwidth]{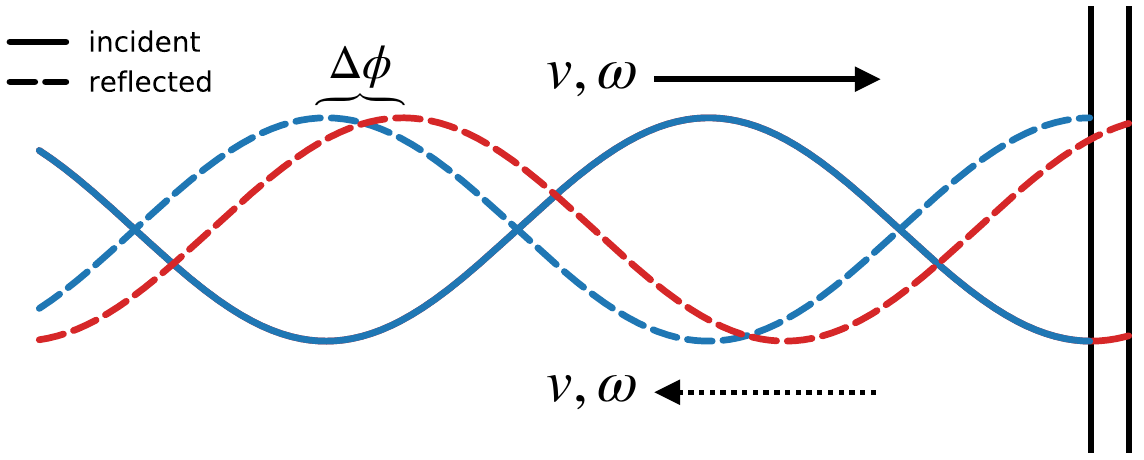}}
  \caption{\emph{Solid lines}: the incoming wave incident on the base
    of the convection zone (vertical black lines); \emph{dashed
      lines}: the waves reflected from the base of the convection
    zone. The blue curves represent the unperturbed case (reflection
    from leftmost vertical line) while the red curves show the effect
    of moving the base of the convection zone to the right (rightmost
    vertical line). The phase difference of the maxima of the
    reflected waves in these two cases is labeled as $\Delta \phi$.  }
  \label{ref1}
\end{figure}

\section{Asymptotic Theory}

We first re-derive a standard result.  From \citet{Gough93} we have
that the asymptotic radial wavenumber is given by
\begin{equation}
  K^2(r) = \frac{\omega^2-\omega_c^2}{c^2} -\frac{L^2}{r^2} \left( 1 -
    \frac{N^2}{\omega^2}\right),
\end{equation}
where $N$ is the \bvfreq, $L^2 \equiv \ell (\ell+1)$, $c$ is the sound
speed, $\omega$ is the angular frequency of the mode, and $\omega_c$
is the acoustic cutoff frequency. The group velocity, $v_{\rm gr}$,
of gravity waves is given by
\citep[see ][]{Unno89,Gough93} 
\begin{eqnarray} 
  v_{\rm gr}  & \equiv & \frac{\partial \omega}{\partial K} = \left(
    \frac{\partial K}{\partial \omega} \right)^{-1}\\
  & = & -\frac{1}{\omega K}\left( L^2 N^2/r^2 \omega^2  - \omega^2/c^2
        \right).
\end{eqnarray} 
For a mode of overtone number $n$, the time taken for it to propagate
from the inner to the outer turning point and back again is
$T_{\rm bounce} = n\, P$. Denoting the inner and outer turning points
as $r_1$ and $r_2$, respectively, we find that
\begin{equation}
  n\,P = T_{\rm bounce}= 2 \int_{r_1}^{r_2} \frac{dr}{|v_{\rm gr}|}.
\label{asymp1}
\end{equation}
In the low-frequency, g-mode limit, $K \approx L N/\omega r$, 
so eqn.~\ref{asymp1} can be written as
\begin{equation}
n \pi = \int_{r_1}^{r_2} K dr.
\label{asymp2}
\end{equation}
This is just the well known quantization condition for asymptotic
nonradial modes.  Finally, the asymptotic formula for mode periods can
be written as
\begin{equation}
  P_n = \frac{2 \pi^2 n}{\sqrt{\ell (\ell+1)}} 
  \left[ \int_{r_1}^{r_2} 
  \frac{N}{r} dr \right]^{-1}.
  \label{gmodes}
\end{equation}

\begin{figure}[t!]
  \centerline{\includegraphics[width=1.0\columnwidth]{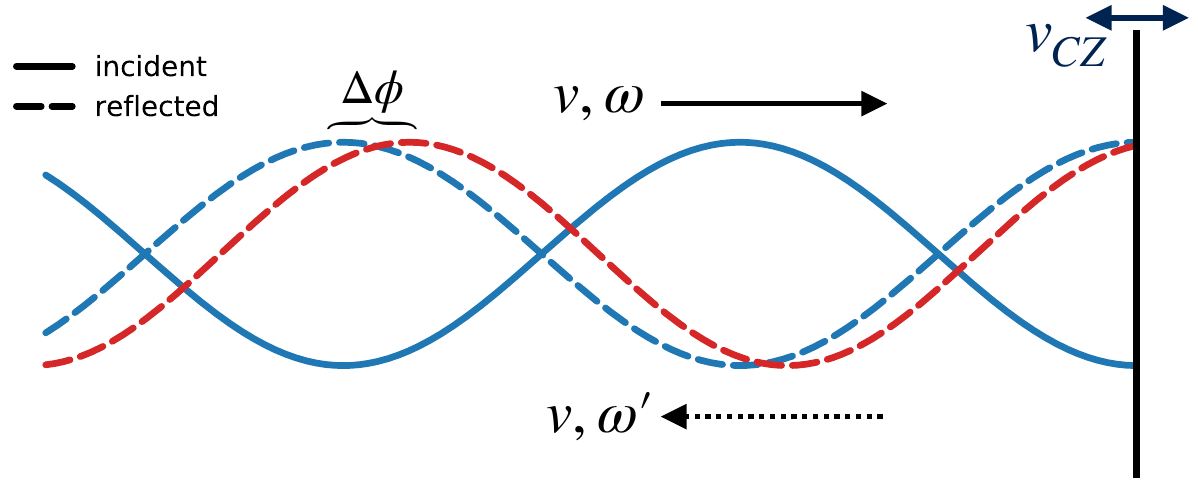}}
  \caption{The change in the frequency of the reflected wave
    ($\omega \rightarrow \omega'$) due to the velocity of the base of
    the convection zone ($v_{CZ}$). This frequency change leads to a
    change in the radial wavenumber of the wave, leading to a slow
    accumulation of phase difference as the wave propagates.  }
  \label{ref2}
\end{figure}

\begin{figure*}[t!]
  \centerline{\includegraphics[width=1.0\textwidth]{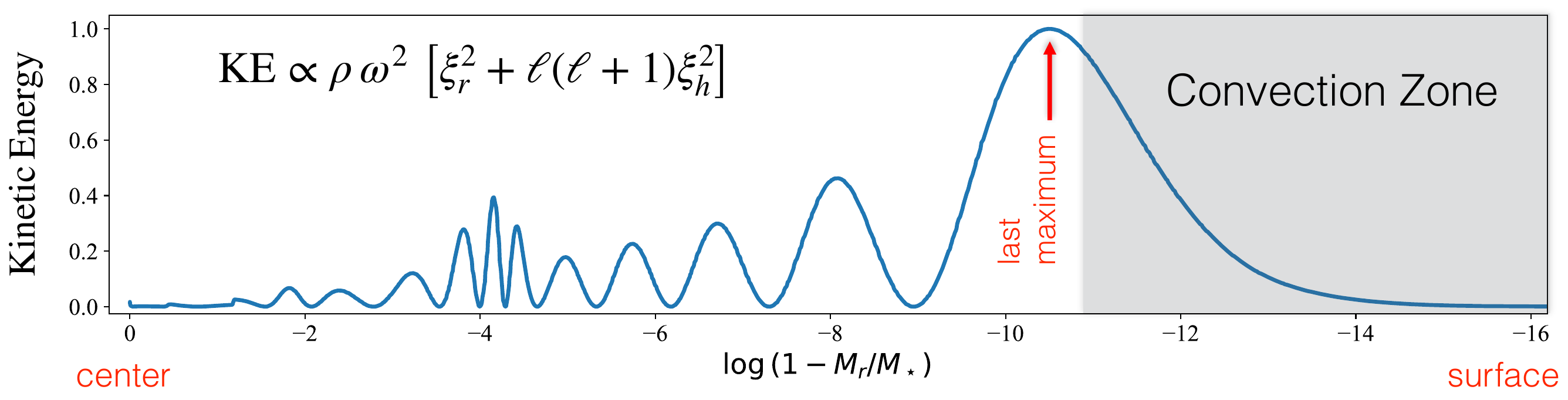}}
  \caption{The position of the last maximum of the kinetic energy
    density (red arrow) before the mode reaches the base of the convection zone.
    At this point, $v_{\rm gr}$ is finite. }
  \label{eig}
\end{figure*}

\section{Interaction with the \\ Convection Zone}
\label{interaction}

There are two main ways that a mode's interaction with the outer
convection zone could lead to a loss of its coherence. First, if a
mode has a low enough frequency, its outer turning point will be the
base of the convection zone.  Since the base of the convection zone
rises and falls with the surface temperature perturbations of the
pulsations, $r_2$ in eqn.~\ref{gmodes} is time dependent. This leads
to variations in $P_n$, which in turn lead to an overall lack of
coherence of the mode (see Fig.~\ref{ref1}). In \citet{Montgomery15b},
we showed that the extra phase of the mode, $\Delta \phi$, could be
related to the variation in period due to this effect by
\begin{equation}
  \Delta \phi_{\rm cav} = 2 \pi n \left(\frac{\Delta P}{P}\right).
  \label{cav}
\end{equation}
This, in turn, will lead to a damping rate for the mode, as we show in
section~\ref{res}.

A second effect that may play an even larger role is the doppler shift
of the wave caused by the motion of the base of the convection zone
(see Fig.~\ref{ref2}). Formally, this is given by
\begin{equation}
  \omega' = \omega \, \left( 1 - v_{CZ}/v_{\rm gr} \right),
  \label{dop}
\end{equation}
where $\omega$ is the frequency of the incident wave, $v_{CZ}$ is the
velocity of the base of the convection zone, and $v_{\rm gr}$ is the
group velocity of the wave at the base of the convection zone. This
frequency difference, as the ray propagates down to the inner turning
point and back out to the outer turning point, results in an
accumulated phase change of
\begin{eqnarray}
\Delta \phi_{\rm dop} & = & 2 \pi n \,\frac{\Delta \omega}{\omega} \\
  & = & -2 \pi n \,\frac{v_{CZ}}{v_{\rm gr}}.
  \label{dop2}
\end{eqnarray}

Equation~\ref{dop2} is non-trivial to evaluate since $v_{\rm gr}$
formally goes to zero at the base of the convection zone. However, it
is possible to get a finite estimate by looking at its effect on the
motion of a well-defined point in the full numerical solution. For
example, by using models that differ only in the depth of their
convection zones, we can calculate how the position of last maximum of
the kinetic energy density (see Fig.~\ref{eig}) changes as a function
of convection zone depth. By treating this point as the effective
reflecting surface, we can compute its velocity ($v_{CZ}$), which,
together with the value of $v_{\rm gr}$ at this point, gives the
doppler shift of the reflected wave.

\section{Theoretical Damping Rates}

For the mode to be completely coherent, it needs to accumulate exactly
$2 \pi n$ radians of phase each time it propagates back and forth in
the star. As shown in the previous section, a changing convection zone
can upset this condition, leading to (slightly) destructive
interference and a change in amplitude given by
\begin{equation}
  \frac{dA}{dt} = - \frac{A}{n P} \left(1-\cos \Delta \phi\right)
\end{equation}
\citep[see][]{Montgomery15b}. Assuming that
$A \propto e^{-\gamma\, t}$, and averaging $\gamma$ over values of the
phase shift from $-\Delta \phi$ to $+\Delta \phi$ gives
\begin{equation}
  \gamma =\frac{1}{n P} \left( 1 -
    \frac{\sin \Delta \phi}{\Delta \phi} \right).
  \label{gamma}
\end{equation}
This is the equation we use to calculate the damping rates of modes
given the amplitudes ($\Delta \phi$) of their phase shifts.

\section{Results}
\label{res}

\begin{figure*}[t!]
  \centerline{
    \includegraphics[width=1.0\textwidth]{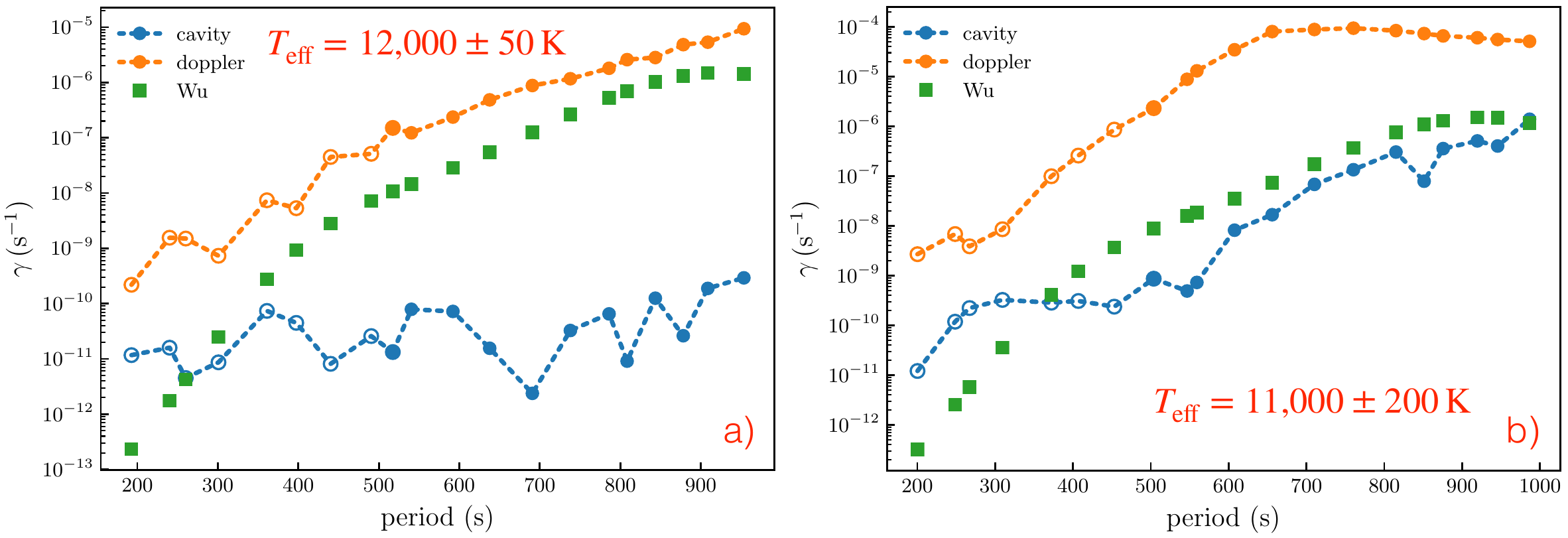}
  }
  \caption{ A comparison of the finite-amplitude damping rates for
    $\ell=1$ modes due to a changing cavity size (blue points,
    eqn.~\ref{cav}) and those due to doppler shifting of the reflected
    wave (orange points, eqn.~\ref{dop2}). We also show an estimate of
    the \emph{linear growth} rate of these modes (green points) based
    on calculations by \citet{Wu98} and \citet{Wu99}. Both sets of
    models have a mass of $0.6\,M_\odot$: those in a) have a
    central \teff\ of 12,000~K with an assumed pulsation amplitude of
    50~K, while those in b) have a central \teff\ of 11,000~K with an
    assumed amplitude of 200~K.  }
  \label{damp}
\end{figure*}

Combining eqns.~\ref{cav} and \ref{dop2} with eqn.~\ref{gamma}, we can
compute finite-amplitude damping rates for different pulsation
modes. In Figs.~\ref{damp}a,b we show the damping rates for $\ell=1$
modes using two different sets of $0.6\,M_\odot$ models. In
Fig.~\ref{damp}a, the models have a central \teff\ of 12,000~K with
assumed temperature excursions of $\pm 50$~K; the observed fractional
flux amplitude would be approximately $\delta F/F \sim 1$\%. The blue
points show the damping rate due to the changing size of the g-mode
cavity, while the orange points give the damping rate due to the
doppler shifting of the reflected wave's frequency. Fig.~\ref{damp}b
is the same but for models with a central \teff\ of 11,000~K with
assumed temperature excursions of $\pm 200$~K; the observed fractional
flux amplitude in this case would be $\delta F/F \sim 4$\%. The green
points in both plots show an estimate of the \emph{linear growth} rate
of these modes based on calculations by \citet{Wu98} and \citet{Wu99}.

\begin{figure}[b!]
  \centering{
  \includegraphics[width=1.0\columnwidth]{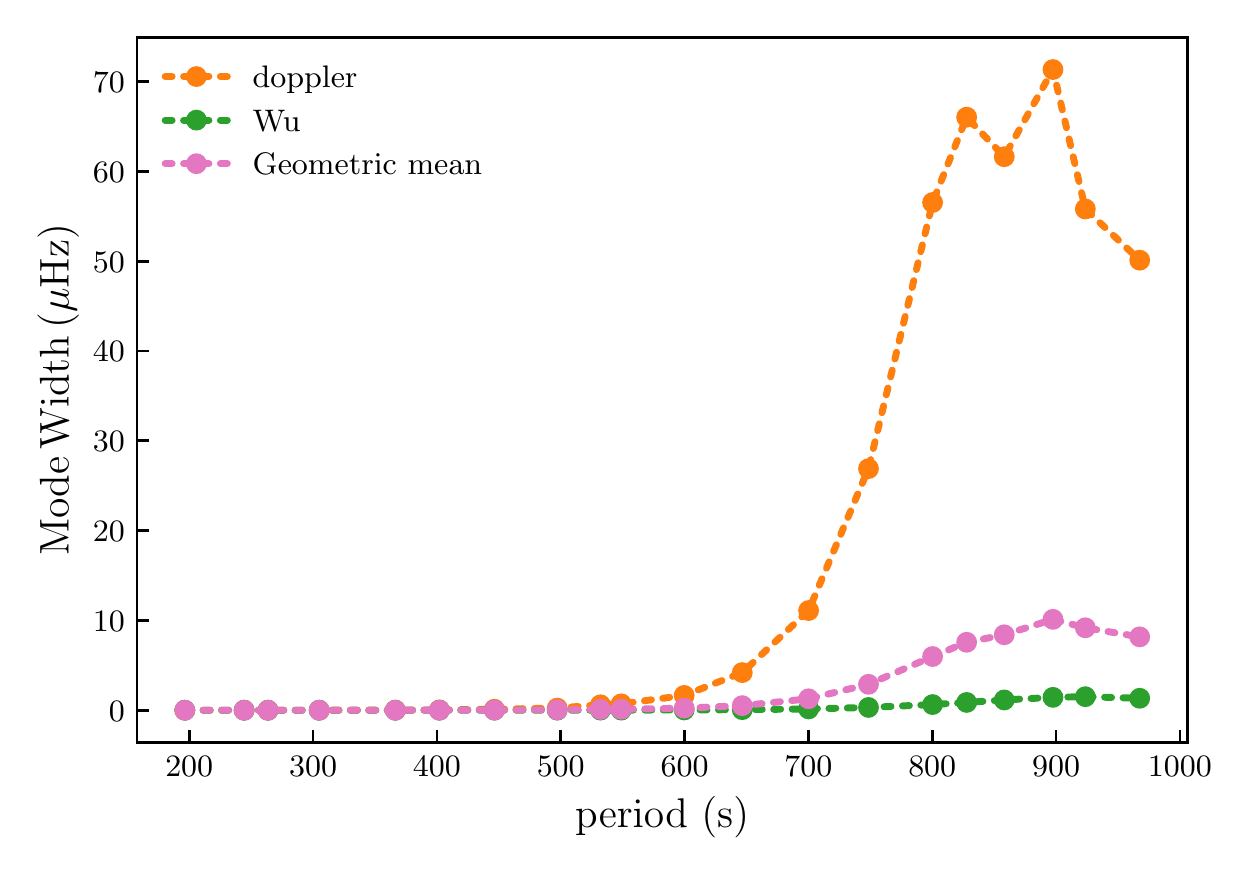}
  }
  \caption{
    Calculation of mode width assuming that the width is equal to the
    finite-amplitude damping (``doppler'', orange points) or linear
    driving (``Wu'', green points) rates. The geometric mean of these
    two calculations is also shown (magenta points).
    }
    \label{widths}
\end{figure}

\begin{figure}[b!]
  \centering{
  \includegraphics[width=0.99\columnwidth]{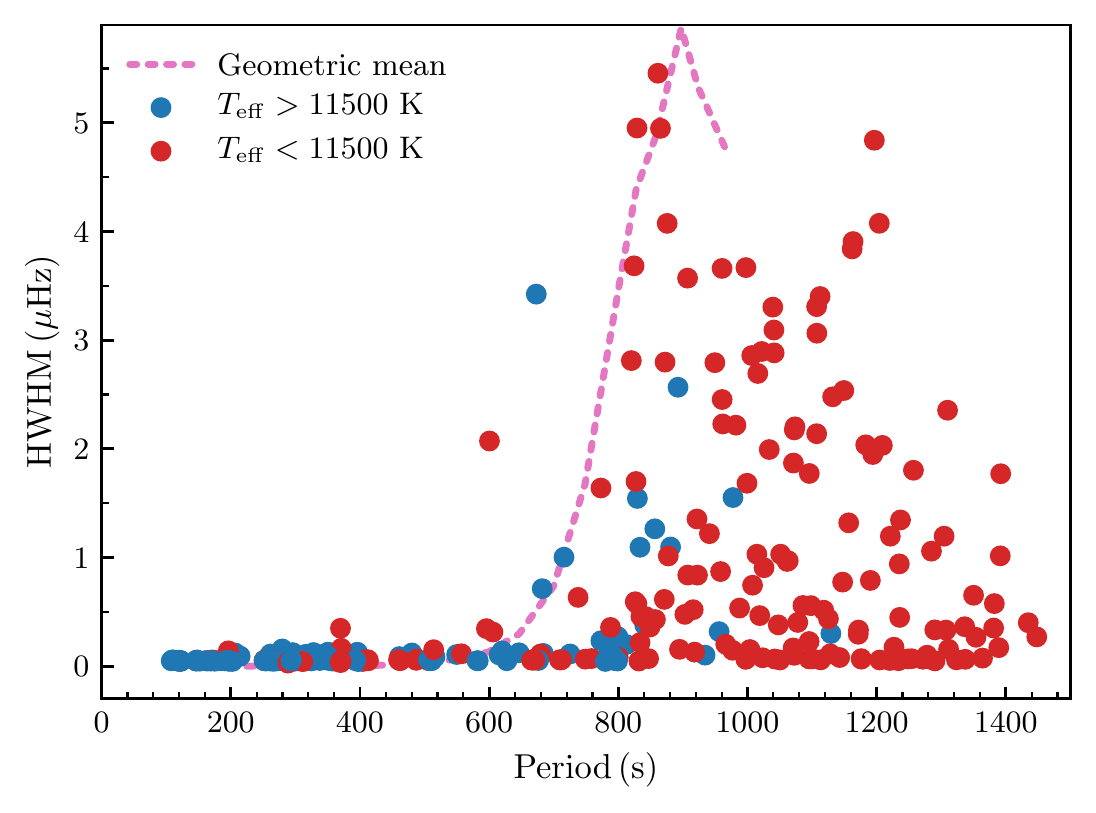}
  }
  \caption{ Comparison of the scaled ``geometric mean'' calculation in
    Fig.~\ref{widths} to the observed mode widths of
    Fig.~\ref{hermes}.  While suggestive that there might be a
    connection between the damping effects considered in
    section~\ref{interaction} and the observed mode widths, further
    work is needed to establish this conclusively.  }
    \label{comp2}
\end{figure}

The linear growth rates, by definition, do not depend on the amplitude
of the pulsations, while the nonlinear damping mechanisms presented
here do. First of all, we see that the damping due to the doppler
shift of the mode's frequency is much larger than that due to the
variation in the size of its g-mode cavity, for all models and modes.
In Fig.~\ref{damp}a, the damping is slightly larger than the driving,
indicating that the predicted \teff\ excursions would be slightly less
than $\pm 50$~K.  In Fig.~\ref{damp}b, the damping is much larger than
the driving, indicating that the predicted \teff\ excursions should be
much smaller than $\pm 200$~K.

Due to the nature of the stochastic excitation mechanism in solar-like
pulsators, the theoretical damping rates provide a prediction for the
observed Fourier widths of the modes. While this connection is less
clear for pulsators with modes that are linearly unstable, we can
still compare the calculated damping and driving rates to the observed
widths of these modes.\footnote{A mode that is linearly driven is
  usually assumed to have grown in amplitude to the point that further
  growth is limited by some nonlinear process, leading to a stable
  limit cycle. At this point its amplitude and phase are nearly
  constant in time, resulting in mode widths that are \emph{much}
  smaller than those given by the calculated driving and damping
  rates.}

Given this assumption, in Fig.~\ref{widths} we plot the mode width due
to damping from the doppler effect (orange curve and points) and that
due to the driving from the Wu and Goldreich convective driving
mechanism (green curve and points). For purely phenomenological
reasons, we also plot the geometric mean of these two quantities
(magenta curve and points).

Finally, in Fig.~\ref{comp2} we plot a scaled version of the geometric
mean from Fig.~\ref{widths} along with the measured mode widths of
\citet{Hermes17a}. We see that the calculation reproduces the
qualitative shape of the envelope of the observed mode widths as a
function of period. This suggests that there might be a connection
between the damping effects considered in section~\ref{interaction}
and the observed mode widths. More work will be needed to establish
this relationship more conclusively.

\vspace*{-1em}
\section{Discussion}
\vspace*{-0.5em}

In the preceding sections we have shown that the modes with observed
Fourier widths greater than 0.3~$\mu$Hz have longer periods
($P \gtrsim 800$~s) and are expected to propagate all the way to the
base of the surface convection zone in pulsating DA WDs. We have put
forward mechanisms through which the convection zone can cause a lack
of coherence in these modes. Preliminary results are that this
mechanism is stronger at the red edge of the instability
strip. However, the size of the effect is such that it may be too
large to be consistent with the observed large pulsation amplitudes (a
few percent) that are seen in these stars.

An additional effect that is surely present at some level is the
overshooting of turbulent fluid motions at the base of the convection
zone.  These motions will produce an effective turbulent viscosity
that will tend to damp the $g$-mode pulsations. This damping, as well
as its variation with time, could also lead to a broadening of the
modes in frequency space. It is also possible that this overshoot
region could at least partially reflect the modes before they have a
chance to reach the base of the convection zone. This could reduce the
size of the damping due to the doppler shift of the reflected waves
that is found in these calculations.


\vspace*{-0.5em}
\section{Conclusions}
\vspace*{-0.5em}

In this paper we present a mechanism that could be relevant to the
properties of white dwarfs as they cool and approach the red edge of
the DAV instability strip. As the convection zone changes its depth
during the pulsation cycle, the condition for coherent reflection of the
outgoing traveling wave is slightly violated. In effect, this causes
the amplitude present in the mode at its linear frequency to slowly
spread to nearby frequencies. This decreases the overall amplitude of
the mode and leads to damping. This mechanism should be present at
some level in all pulsating WDs, and should be larger near the red
edge of the DAV instability strip. Preliminary calculations show that
this damping is probably too large to be consistent with the observed
amplitudes in some DAVs, although other effects may operate which
mitigate this.  In addition, this mechanism could possibly be relevant
for other g-mode pulsators with surface convection zones (e.g., Gamma
Doradus stars), or even large-amplitude pulsators such as RR Lyrae
stars or high-amplitude Delta Scuti stars (HADs).

\vspace*{-1em}

\vspace*{-0.5em}
\section*{Acknowledgements}
\vspace*{-0.5em} MHM and DEW acknowledge support from the United
States Department of Energy under grant DE-SC0010623, the Wootton
Center for Astrophysical Plasma Properties under the United States
Department of Energy under grant DE-FOA-0001634, and the NSF grant AST
1707419. JJH acknowledges support from NASA through Hubble Fellowship
grant \#HST-HF2-51357.001-A, awarded by the Space Telescope Science
Institute, which is operated by the Association of Universities for
Research in Astronomy, Incorporated, under NASA contract NAS5-26555.



\bibliography{../../../styles/index_f.bib}



\end{document}